\documentclass[paper,11pt]{article}
\pdfoutput=1

\usepackage{jheppub}

\usepackage[T1]{fontenc}
\usepackage{amsmath}
\usepackage{mdframed}
\usepackage{graphics}
\usepackage{amsfonts}
\usepackage{amssymb}
\usepackage{bm}
\usepackage{amsthm}
\usepackage{slashed}
\usepackage{enumerate,enumitem}
\usepackage{amsbsy}
\usepackage{bbm}
\usepackage{mathrsfs}
\usepackage{url}
\usepackage{bbm}
\usepackage{color}
\usepackage{framed}
\usepackage{subfig}
\usepackage{caption}
\usepackage{float}
\usepackage{cancel}
\usepackage{hyperref}
\usepackage{comment}

\newcommand{\tr}{\mathrm{tr}}
\newcommand{\R}{\mathbb{R}}
\newcommand{\cA}{\mathcal{A}}
\newcommand{\cS}{\mathcal{S}}
\newcommand{\cV}{\mathcal{V}}

\newcommand{\cO}{\mathcal{O}}

\newcommand{\cL}{\mathcal{L}}
\newcommand{\bcL}{\bar{\mathcal{L}}}

\newcommand{\Bx}{\mathbf{x}}

\newcommand{\BT}{\mathbf{T}}

\newcommand{\BD}{\mathbf{D}}
\newcommand{\BP}{\mathbf{P}}
\newcommand{\BG}{\mathbf{G}}
\newcommand{\BF}{\mathbf{F}}
\newcommand{\BH}{\mathbf{H}}

\newcommand{\dd}{\mathrm{d}}

\newcommand{\bS}{\bar{S}}
\newcommand{\bP}{\bar{P}}
\newcommand{\ta}{\tau}
\newcommand{\bta}{\bar\tau}
\newcommand{\TbT}{\mathrm{T}\overline{\mathrm{T}}}

\newcommand{\hT}{\widehat{T}}
\newcommand{\T}{T}
\newcommand{\Id}{\mathbbm{1}}
\newcommand{\hg}{\hat{g}}
\newcommand{\cg}{\check{g}}
\newcommand{\tF}{\widetilde{F}}
\newcommand{\var}{s}

\newcommand{\D}{d}
\newcommand{\C}{c}

\newcommand{\Sc}{\scriptscriptstyle{\text{S}}}
\newcommand{\M}{\scriptscriptstyle{\text{M}}}
\newcommand{\MBI}{\scriptscriptstyle{\text{MBI}}}
\newcommand{\MM}{\scriptscriptstyle{\text{MM}}}
\newcommand{\MMBI}{\scriptscriptstyle{\text{MMBI}}}
\newcommand{\MS}{\scriptscriptstyle{\text{MS}}}
\newcommand{\MSBI}{\scriptscriptstyle{\text{MSBI}}}

\title{Metric approach to a $\mathrm{T}\overline{\mathrm{T}}-$like deformation in arbitrary dimensions}

\author[1]{Riccardo Conti,}
\author[2]{Jacopo Romano,}
\author[3]{Roberto Tateo}

\affiliation[1]{Departamento de Matem\'atica, Faculdade de Ci\^encias da Universidade de
Lisboa, Campo Grande Edif\'icio C6, 1749-016, Lisboa, Portugal.}

\affiliation[2]{Max Planck Institute for Dynamics and Self-Organization, Am Faßberg 17, 37077, Gottingen, Germany.}

\affiliation[3]{Dipartimento di Fisica and Arnold-Regge Center, Universit\`a di Torino, and INFN Sezione di Torino, Via P. Giuria 1, 10125, Torino, Italy.}

\emailAdd{rconti@fc.ul.pt}
\emailAdd{jacopo.romano@ds.mpg.de}
\emailAdd{roberto.tateo@unito.it}
  
\abstract{
We consider a one-parameter family of composite fields -- bi-linear in the components of the stress-energy tensor -- which generalise the $\mathrm{T}\overline{\mathrm{T}}$ operator to arbitrary space-time dimension $d\geq 2$. We show that they induce a deformation of the classical action which is equivalent -- at the level of the dynamics -- to a field-dependent modification of the background metric tensor according to a specific flow equation. Even though the starting point is the flat space, the deformed metric is generally curved for any $d>2$, thus implying that the corresponding deformation can not be interpreted as a coordinate transformation. The central part of the paper is devoted to the development of a recursive algorithm to compute the coefficients of the power series expansion of the solution to the metric flow equation. We show that, under some quite restrictive assumptions on the stress-energy tensor, the power series yields an exact solution. Finally, we consider a class of theories in $d=4$ whose stress-energy tensor fulfils the assumptions above mentioned, namely the family of abelian gauge theories in $d=4$. For such theories, we obtain the exact expression of the deformed metric and the vierbein. In particular, the latter result implies that ModMax theory in a specific curved space is dynamically equivalent to its Born-Infeld-like extension in flat space. We also discuss a dimensional reduction of the latter theories from $d=4$ to $d=2$ in which an interesting marginal deformation of $d=2$ field theories emerges.
}

\begin{document}
\maketitle
\flushbottom

\section{Introduction}
\label{sec:intro}
The recent discovery  that specific irrelevant perturbations \cite{Zamolodchikov:2004ce} of field theories in dimension $\D=2$ can be addressed using exact flow equations  \cite{Smirnov:2016lqw, Cavaglia:2016oda}  and other powerful mathematical tools \cite{Caselle:2013dra, Dubovsky:2012wk, Cardy:2018sdv}, has triggered a fair amount of research activity. 
Particularly striking are the observed links with string theory \cite{Cavaglia:2016oda} and  topological gravity \cite{Dubovsky:2017cnj}, together with the AdS/CFT interpretation of these perturbations \cite{McGough:2016lol}. The main motivations to study these novel class of models are the deepening of our general knowledge on non-renormalisable Quantum Field Theories and to clarify aspects of quantum gravity.

In this paper, we work within the framework of Lagrangian field theories in space-time dimension $\D\geq 2$ equipped with a metric tensor $g_{\mu\nu}=g_{\mu\nu}(\Bx)$ with Euclidean signature, where $\Bx=(x^0,x^1,\dots,x^{\D-1})$ is a set of local coordinates. We denote as
\begin{align}
\label{eq:covaction}
&\cA=\int \dd^\D\Bx\sqrt{g}\cL = \int \dd^\D\Bx\,\bcL \;,\\
&g:=\det[g_{\mu\nu}] \;,\quad \dd^\D\Bx:=\dd x^0\,\dd x^1\dots\dd x^{\D-1}\;,\notag
\end{align}
a generic covariant action where $\bcL := \sqrt{g}\, \cL$ is the Lagrangian density that depends on $\Bx$ through a generic collection of $N$ fields $\lbrace  \Phi_I\rbrace_{I\in\{1,\dots,N\}}$ and their higher-order derivatives $\{\partial_{\mu_1}\dots\partial_{\mu_i}\Phi_I\}_{(I,i)\in\{1,\dots,N\}\times\{1,\dots,n\}}$ for some $n\geq 1$ and with $\partial_{\mu}=\frac{\partial}{\partial x^\mu}$. The field content of the theory is arbitrary, unless otherwise stated. Indexes of tensors are lowered and raised using the metric $g_{\mu\nu}$ and its inverse $g^{\mu\nu}$, respectively, and repeated indexes are summed according to the Einstein notation. We shall denote with $\eta_{ab}$ the flat metric with the same (Euclidean) signature of $g_{\mu\nu}$. Following the standard convention, we use latin (Lorentz) and greek (Einstein) indexes to distinguish between flat and curved reference frames, respectively, and we adopt the tetrad formalism to move from one to the other as customary.

This paper focuses on a family of deformations defined by the flow equation
\begin{equation}
    \dfrac{\partial\cA_\ta}{\partial\ta} = \int \dd^\D\Bx\sqrt{g}\,\cO_\ta^{[r,\D]} \;,\quad
\cA_{\ta_0}=\cA\;,
\label{eq:genericflow}
\end{equation}
with perturbing operator \footnote{ It is important to stress that this paper is about classical field theories and, apart from the special case $(r,d)=(1,2)$ \cite{Zamolodchikov:2004ce},  it is not known how to make the composite field (\ref{eq:Od}) well-defined also at the quantum level.}
\begin{equation}
\label{eq:Od}
    \cO_\ta^{[r,\D]}:=\frac{1}{\D}\left(r\,\tr[\BT_\ta]^2-\tr[\BT_\ta^2]\right)\;,\quad r\in\R\;,\;\D\geq 2\;,
\end{equation}
where $\ta\in\R$ is the flow parameter and $\ta_0$ is a fixed value; $\cA_\ta=\int\dd^\D\Bx\,\bcL_\ta$ denotes the deformed action and $\bcL_\ta$ the corresponding Lagrangian density; $\BT_\ta=\left(\T_{\ta,\nu}^{\mu}\right)_{\mu,\nu\in\{0,\dots,\D-1\}}$ is a $\D\times \D$ matrix and $\T_\ta^{\mu\nu}$ are the components of the (symmetric) Hilbert stress-energy tensor associated to $\cA_\ta$ according to the standard prescription
\begin{equation}
\label{eq:Hstressen}
\T_\ta^{\mu\nu}=\dfrac{-2}{\sqrt{g}}\dfrac{\delta\cA_\ta}{\delta g_{\mu\nu}} = \dfrac{-2}{\sqrt{g}}\dfrac{\partial \bcL_\ta}{\partial g_{\mu\nu}}\;.
\end{equation}
We start by reviewing some facts about the most studied representative among this family of deformations, namely the $\TbT$ deformation of field theories in $\D=2$, from which the present paper draws inspiration.

The $\TbT$ deformation \cite{Smirnov:2016lqw, Cavaglia:2016oda} is described by the flow equation \eqref{eq:genericflow} with $(r,\D)=\left(1,2\right)$, i.e. the $\TbT$ operator is given by
\begin{equation}
\cO_\ta^{\TbT}:=\cO_\ta^{[1,2]}=\frac12\left(\tr[\BT_\ta]^2-\tr[\BT_\ta^2]\right)=\det[\BT_\ta]\;,
\label{eq:OTT}
\end{equation}
where in the last equality we used the Cayley-Hamilton Theorem. The $\TbT-$deformed action $\cA_\ta$ can be obtained either directly by solving explicitly the flow equation \eqref{eq:genericflow} or indirectly using a field-dependent coordinate transformation \cite{Dubovsky:2017cnj,Conti:2018tca} (see also \cite{Caputa:2020lpa}) which provides an efficient tool to derive also solutions of the $\TbT-$deformed equations of motion \cite{Conti:2018tca,Ceschin:2020jto} and integrals of motion \cite{Conti:2019dxg}. Let us also mention that an alternative method to compute $\TbT$-deformed actions is given by the light-cone gauge approach developed in \cite{Frolov:2019nrr, Frolov:2019xzi}.

As it was noted in \cite{Conti:2018tca}, the coordinate transformation induces a specific field-dependent deformed metric that defines a modified background in which the solutions of the seed theory are equivalent to the corresponding $\TbT-$deformed ones in flat space. In other words, there exists a deformed metric that makes the seed theory dynamically equivalent\footnote{Because the equivalence is at the level of the equations of motion.} to the deformed theory in flat space. Strictly speaking, this deformed metric is a pseudo-metric since, for a generic field configuration there might exist a range of values of the deformation parameter for which it becomes degenerate (see \cite{Conti:2018tca}). 

Throughout the paper we will handle with pseudo-metrics -- see \eqref{eq:hg} and \eqref{eq:cg} -- associated to the generalised operators \eqref{eq:Od} in arbitrary dimension $d\geq2$. However, we shall refer to them simply as metrics neglecting the  issue related to the degeneracy, since it does not affect the general conclusions that emerge from our analysis.

In contrast to the $\TbT$ operator, the geometric properties of the operators \eqref{eq:Od} for $\D>2$ are essentially unknown. The interest toward such deformations is partially due to the discovery first made in \cite{Conti:2018jho}, that the operator $\cO_\ta^{\left[\frac12,4\right]}$ surprisingly links the Maxwell theory with Maxwell Born-Infeld \cite{Plebanski:1970zz} and, in \cite{Babaei-Aghbolagh:2022uij,Ferko:2022iru}, it was proven that the same link exists between the ModMax theory \cite{Bandos:2020jsw} and its Born-Infeld-like extension \cite{Bandos:2020hgy}, thus generalising the result of \cite{Conti:2018jho}. 

The aim of this paper is to study the geometric properties of the family of deformations \eqref{eq:genericflow} through a metric approach. In section \ref{sec:metricapproach}, we start by showing that \eqref{eq:genericflow} can be interpreted as a modification of the background metric -- at dynamical level -- according to a specific flow equation. In section \ref{sec:defcurv} we prove that, for a generic field configuration, such deformed metric is curved except for the specific case $(r,\D)=(1,2)$
-- corresponding to the $\TbT$ deformation -- in which it remains flat, in accordance with the existence of a coordinate transformation. In section \ref{sec:metricflow}, we develop a perturbative algorithm to solve the flow equation for the metric and, in section \ref{sec:metrictrunc}, we show that under some assumptions on the stress-energy tensor, the series yields an exact solution for the metric. In section \ref{sec:abelian} we consider the class of abelian gauge theories in $\D=4$, whose stress-energy tensors meet the conditions above-mentioned, and we derive an exact expression for the deformed metric and the vierbein; appendix \ref{sec:appendix} contains the details of the derivation of the vierbein. Finally, in section \ref{sec:MMBIdimred} we construct a class of modified scalar theories in $\D=2$ and their corresponding $\TbT$ deformation, as a dimensional reduction of the ModMax theory and its Born-Infeld-like extension.
\section{A $\TbT-$like deformation in $\D$ dimensions}
\label{sec:4Ddef}
For the purposes of the current paper, it is convenient to rewrite \eqref{eq:Od} as follows
\begin{equation}
    \cO_\ta^{[r,\D]}=\frac{1}{\D}\hT_{\ta,\mu\nu}\T_\ta^{\mu\nu}\;,
\end{equation}
where we introduced the tensor
    \begin{equation}
    \hT_{\ta,\mu\nu}:=f_{\mu\nu\rho\sigma}\T_\ta^{\rho\sigma}=r g_{\mu\nu}\tr[\BT_\ta]-\T_{\ta,\mu\nu}\;,\label{eq:hTdef}
    \end{equation}
    and
    \begin{equation}
        f_{\mu\nu\rho\sigma}:=r\, g_{\mu\nu}g_{\rho\sigma}-g_{\mu\sigma}g_{\nu\rho} \;.
    \end{equation}
It is immediate to check that $f_{\mu\nu\rho\sigma}$ fulfils the following properties
\begin{equation}
f_{\rho\sigma\mu\nu}=f_{\mu\nu\rho\sigma}\;,\quad f_{\nu\mu\sigma\rho}=f_{\mu\nu\rho\sigma}\;,\quad f_{\mu\nu\alpha\beta}f^{\alpha\beta\rho\sigma}=r\left(\D r-2\right)g_{\mu\nu}g^{\rho\sigma}+\delta_\mu^\rho\delta_\nu^\sigma \;.
\label{eq:fprop}
\end{equation}
Notice that $r=0$ and $r=\frac{2}{\D}$ seems to be special cases since the last formula in \eqref{eq:fprop} simplifies.

\subsection{Metric approach}
\label{sec:metricapproach}
In this section, we prove that \eqref{eq:genericflow} amounts to a modification of the background metric at the level of the dynamics and we identify the flow equation that describes the evolution of the metric.
To this aim, we adopt the same logic followed by \cite{Cardy:2018sdv} in the $\TbT$ context. Under an infinitesimal deformation $\delta\ta$ of the parameter $\ta$, \eqref{eq:genericflow} can be written as
\begin{equation}
\label{eq:cAinf}
 \cA_{\ta+\delta \ta}(g_{\mu\nu})= \cA_\ta(g_{\mu\nu})+\delta\ta\int \dd^\D\Bx\sqrt{g}\,\cO_\ta^{[r,\D]}\;,
\end{equation}
where we explicitly reported the dependence of the action on the background metric for future convenience. Let $\delta g_{\mu\nu}=\delta\ta\,h_{\mu\nu}$ be an infinitesimal deformation of the metric where $h_{\mu\nu}$ is dynamical, and consider the action
\begin{align}
\hat\cA(h_{\mu\nu}) &= \cA_\ta(g_{\mu\nu}+\delta \ta h_{\mu\nu})+\C\,\delta\ta\int \dd^\D\Bx\sqrt{g}\,
   e^{\mu \nu\rho \sigma}h_{\mu\nu}h_{\rho\sigma} \notag\\
   &= \cA_\ta(g_{\mu\nu})+\delta\ta\int \dd^\D\Bx\sqrt{g}\left[\C\,e^{\mu \nu\rho \sigma}h_{\mu\nu}h_{\rho\sigma}-\frac12 h_{\mu\nu}\T_\ta^{\mu\nu}\right] \;,
\label{eq:cAh}
\end{align}
where $\C$ is a real constant and we defined the tensor
\begin{equation}
    e_{\mu\nu\rho\sigma}:=q\, g_{\mu\nu}g_{\rho\sigma}-g_{\mu\sigma}g_{\nu\rho} \;,\quad q\in\R\;.
\end{equation}
Notice that $e_{\mu\nu\rho\sigma}$ has the same form as $f_{\mu\nu\rho\sigma}$, thus it fulfils the same properties \eqref{eq:fprop} with the substitution $(f_{\mu\nu\rho\sigma},r)\to(e_{\mu\nu\rho\sigma},q)$. Moreover, the two tensors are trivially related via
\begin{equation}
\label{eq:fVSe}
    e_{\mu\nu\rho\sigma}=f_{\mu\nu\rho\sigma}+\left(q-r\right)g_{\mu\nu}g_{\rho\sigma}\;.
\end{equation}
In the following we will fix the parameters $(\C,q)$ in \eqref{eq:cAh} in terms of $(r,d)$ by requiring that the actions \eqref{eq:cAinf} and \eqref{eq:cAh} are dinamically equivalent, i.e. they have the same equations of motion
\begin{equation}
\label{eq:cAequiv}
\frac{\delta\cA_{\ta+\delta\ta}(g_{\mu\nu})}{\delta\Phi_I}=\left.\frac{\delta\cA_\ta(g_{\mu\nu}+\delta\ta h_{\mu\nu})}{\delta\Phi_I}\right|_{h_{\mu\nu}=h_{\mu\nu}^*} \Longleftrightarrow\; \cA_\ta(g_{\mu\nu}+\delta\ta h_{\mu\nu}^*)\simeq \cA_{\ta+\delta\ta}(g_{\mu\nu})\;,
\end{equation}
where we introduced the symbol $\simeq$ to denote the dynamical equivalence between the actions. Notice that in \eqref{eq:cAequiv} we used the fact that
\begin{equation}
    \frac{\delta\hat\cA(h_{\mu\nu})}{\delta\Phi_I} = \frac{\delta\cA_\ta(g_{\mu\nu}+\delta\ta h_{\mu\nu})}{\delta\Phi_I}\;,
\end{equation}
and the variation of $\cA_\ta(g_{\mu\nu}+\delta\ta h_{\mu\nu})$ w.r.t. $\Phi_I$ is performed before evaluating $h_{\mu\nu}$ to its on shell value $h_{\mu\nu}^*$.

We first compute the variation of $\hat\cA(h_{\mu\nu})$ w.r.t. $h_{\mu\nu}$ and set it to zero to obtain the equation of motion for $h_{\mu\nu}$. Using the analogous of the properties \eqref{eq:fprop} in which $(f_{\mu\nu\rho\sigma},r)\to(e_{\mu\nu\rho\sigma},q)$ and formula \eqref{eq:fVSe} we have
\begin{align}
\dfrac{\delta\hat\cA(h_{\mu\nu})}{\delta h_{\mu\nu}}=0 \;&\Longleftrightarrow\; 2\C\,e^{\mu\nu\rho\sigma}h_{\rho\sigma}-\frac12 \T_\ta^{\mu\nu}=0\;\Longleftrightarrow\;2\C\,e_{\mu\nu\alpha\beta}e^{\alpha\beta\rho\sigma}h_{\rho\sigma}-\frac12e_{\mu\nu\rho\sigma}\T_\ta^{\rho\sigma}=0\notag\\
&\Longleftrightarrow\; 2\C\left[h_{\mu\nu}+q\left(dq-2\right)g_{\mu\nu}g^{\rho\sigma}h_{\rho\sigma}\right]-\frac12\hT_{\ta,\mu\nu}+\frac{r-q}{2}g_{\mu\nu}\tr[\BT_\ta]=0\;.
\label{eq:hEoM}
\end{align}
Multiplying both sides of the last equation in \eqref{eq:hEoM} by $g^{\mu\nu}$ we obtain
\begin{equation}
\label{eq:htr}
g^{\mu\nu}h_{\mu\nu}=\frac{\tr[\BT_\ta]}{4\C\left(dq-1\right)}\;.
\end{equation}
Finally, plugging \eqref{eq:htr} back into the last equation in \eqref{eq:hEoM} we can write the equation of motion for $h_{\mu\nu}$ as $h_{\mu\nu}=h_{\mu\nu}^*$ with
\begin{equation}
\label{eq:hsol}
h_{\mu\nu}^*:=\frac{1}{4\C}\left[\hT_{\ta,\mu\nu}+\left(\frac{q}{dq-1}-r\right)g_{\mu\nu}\tr[\BT_\ta]\right]\;.
\end{equation}
The next step is to compute the variations of both $\hat\cA(h_{\mu\nu})$ and $\cA_{\ta+\delta\ta}(g_{\mu\nu})$ w.r.t. $\Phi_I$. For $\hat\cA(h_{\mu\nu})$ we have
\begin{align}
\label{eq:PhiEoMhcA}
    \frac{\delta\hat\cA(h_{\mu\nu})}{\delta\Phi_I} = \frac{\delta\cA_\ta(g_{\mu\nu}+\delta\ta h_{\mu\nu})}{\delta\Phi_I}&=
    \frac{\delta\cA_{\ta}(g_{\mu\nu})}{\delta\Phi_I}-\frac12\delta\ta\Biggl[\sqrt{g}\,h_{\rho\sigma}     \dfrac{\partial\T_\ta^{\rho\sigma}}{\partial\Phi_I}\notag\\
    &+\sum_{i=1}^n(-1)^i\partial_{\mu_1}\dots\partial_{\mu_i}\left(\sqrt{g}\,h_{\rho\sigma}\frac{\partial\T_\ta^{\rho\sigma}}{\partial(\partial_{\mu_1}\dots\partial_{\mu_i}\Phi_I)}\right)\Biggr]\;,
\end{align}
while for $\cA_{\ta+\delta\ta}(g_{\mu\nu})$ we have
\begin{align}
\label{eq:PhiEoMcA}
    &\frac{\delta\cA_{\ta+\delta\ta}(g_{\mu\nu})}{\delta\Phi_I}=\frac{\delta\cA_{\ta}(g_{\mu\nu})}{\delta\Phi_I}+\delta\ta\Biggl[\sqrt{g}\,\dfrac{\partial\cO_\ta^{[r,d]}}{\partial\Phi_I}+\sum_{i=1}^n(-1)^i\partial_{\mu_1}\dots\partial_{\mu_i}\left(\sqrt{g}\,\frac{\partial\cO_\ta^{[r,d]}}{\partial(\partial_{\mu_1}\dots\partial_{\mu_i}\Phi_I)}\right)\Biggr]\notag\\
    &=\frac{\delta\cA_{\ta}(g_{\mu\nu})}{\delta\Phi_I}+\frac{2\delta\ta}{d}\Biggl[\sqrt{g}\,\hT_{\ta,\mu\nu}\frac{\partial\T_\ta^{\mu\nu}}{\partial\Phi_I}+\sum_{i=1}^n(-1)^i\partial_{\mu_1}\dots\partial_{\mu_i}\left(\sqrt{g}\,\hT_{\ta,\rho\sigma}\frac{\partial\T_\ta^{\rho\sigma}}{\partial(\partial_{\mu_1}\dots\partial_{\mu_i}\Phi_I)}\right)\Biggr]\;.
\end{align}
Notice that in the last equality of \eqref{eq:PhiEoMcA} we used the fact that
\begin{align}
   \dfrac{\partial\cO_\ta^{[r,d]}}{\partial X}&=\frac{1}{d}\left(\hT_{\ta,\mu\nu}\dfrac{\partial\T_\ta^{\mu\nu}}{\partial X}+\T_\ta^{\mu\nu}\dfrac{\partial\hT_{\ta,\mu\nu}}{\partial X}\right)=\frac{1}{d}\left[\hT_{\ta,\mu\nu}\dfrac{\partial\T_\ta^{\mu\nu}}{\partial X}+\T_\ta^{\mu\nu}\left(r\,g_{\mu\nu}\frac{\partial\,\tr[\BT_\ta]}{\partial X}-\frac{\partial\T_{\ta,\mu\nu}}{\partial X}\right)\right]\notag\\
   &=\frac{1}{d}\left[\hT_{\ta,\mu\nu}\dfrac{\partial\T_\ta^{\mu\nu}}{\partial X}+\T_\ta^{\mu\nu}\left(r\,g_{\mu\nu}g_{\rho\sigma}\frac{\partial\T_\ta^{\rho\sigma}}{\partial X}-\frac{\partial\T_{\ta,\mu\nu}}{\partial X}\right)\right]\notag\\
   &=\frac{1}{d}\left[\hT_{\ta,\mu\nu}\dfrac{\partial\T_\ta^{\mu\nu}}{\partial X}+\left(r\,g_{\mu\nu}\tr[\BT_\ta]-\T_{\ta,\mu\nu}\right)\frac{\partial\T_\ta^{\mu\nu}}{\partial X}\right]  =\frac{2}{d}\hT_{\ta,\mu\nu}\frac{\partial\T_\ta^{\mu\nu}}{\partial X}\;,
\end{align}
where $X$ is any element of the set $\{\partial_{\mu_1}\dots\partial_{\mu_i}\Phi_I\}_{(I,i)\in\{1,\dots,N\}\times\{1,\dots,n\}}$.
From \eqref{eq:hsol}, \eqref{eq:PhiEoMhcA} and \eqref{eq:PhiEoMcA} it is immediate to see that the equivalence \eqref{eq:cAequiv} holds, in general, only if the parameters $(\C,q)$ are chosen as follows
\begin{equation}
\label{eq:equivcond}
(\C,q)=\left(-\frac{\D}{16},\frac{r}{dr-1}\right)
\quad\Longrightarrow \quad h_{\mu\nu}^*=-\frac{4}{\D}\hT_{\ta,\mu\nu}\;.
\end{equation}
In the following, we shall impose the constraint \eqref{eq:equivcond}. Using the identity
\begin{equation}
\cA_\ta(g_{\mu\nu}+\delta\ta h_{\mu\nu})=\cA_\ta\left(g_{\mu\nu}(\ta+\delta\ta)-\delta\ta\left[\frac{\dd g_{\mu\nu}}{\dd\ta}-h_{\mu\nu}\right]\right)\;,
\end{equation}
in \eqref{eq:cAequiv}, we obtain the following (constrained) dynamical equivalence
\begin{equation}
\label{eq:cAequivfin}
\begin{cases}
\cA_{\ta+\delta\ta}(g_{\mu\nu})\simeq\cA_\ta(g_{\mu\nu}(\ta+\delta\ta))\\
\displaystyle{\frac{\dd g_{\mu\nu}}{\dd\ta}=-\frac{4}{\D}\hT_{\ta,\mu\nu}}
\end{cases}\;,
\end{equation}
which has the following physical interpretation: the deformed theory $\cA_{\ta+\delta\ta}$ with background metric $g_{\mu\nu}(\ta)$ is dynamically equivalent to the theory $\cA_\ta$ with deformed background metric $g_{\mu\nu}(\ta+\delta\ta)$, which evolves according to the second equation of \eqref{eq:cAequivfin}.

Notice that \eqref{eq:cAequivfin} can be equivalently written as
\begin{equation}
\label{eq:cAequivfin2}
\begin{cases}
    \cA_\ta(g_{\mu\nu})\simeq \cA_{\ta+\delta\ta}(g_{\mu\nu}(\ta+\delta\ta))\\ \displaystyle{\dfrac{\dd g_{\mu\nu}}{\dd\ta}=\frac{4}{\D}\hT_{\ta,\mu\nu}}
    \end{cases}\;,
\end{equation}
which has the following physical interpretation: the theory $\cA_\ta$ with background metric $g_{\mu\nu}(\ta)$ is dynamically equivalent to the deformed theory $\cA_{\ta+\delta\ta}$ with deformed background metric $g_{\mu\nu}(\ta+\delta\ta)$, which evolves according to the second equation of \eqref{eq:cAequivfin2}.

\subsection{Deformation of the Riemann tensor}
\label{sec:defcurv}
In this section we briefly discuss the infinitesimal deformation of the Riemann tensor $\delta R_{\sigma\mu\nu}^\rho$ induced by the infinitesimal deformation $\delta g_{\mu\nu}=-\frac{4}{\D}\delta\ta\,\hT_{\ta,\mu\nu}$ of the metric. Assuming that the starting point is a $\D-$dimensional flat space with metric $\eta_{ab}$, i.e. the associated Riemann tensor is $R^i_{jab}= 0$, then $\delta g_{\mu\nu}:=\delta\eta_{ab}e^a_\mu e^b_\nu$ where we defined $\delta \eta_{ab}=-\frac{4}{\D}\delta\ta\,\hT_{\ta,ab}$ and $e^a_\mu=\delta^a_\mu$ is the trivial vierbein.

A standard computation leads to
\begin{equation}
\delta R_{jab}^i=\frac{2}{\D}\delta\ta\left(\partial_b \partial_j \hT_{\ta,a}^i-\partial_a\partial_j \hT_{\ta,b}^i+\partial_a \partial^i \hT_{\ta,jb}-\partial_b \partial^i \hT_{\ta,ja} \right)\;,
\label{eq:deltaRiem}
\end{equation}
for the Riemann tensor,
\begin{equation}
\delta R_{ab}=\delta R^i_{aib}=\frac{2}{\D}\delta\ta   \left(\partial_i\partial^i \hT_{\ta,ab}+\left(r\D-2r-1\right)
\partial_a\partial_b\tr[\BT_\ta]\right)\;,
\label{eq:deltaRic}
\end{equation}
for the Ricci tensor and
\begin{equation}
    \delta R=R_{ab}\,\delta \eta^{ab}+\eta^{ab}\delta R_{ab}=
\frac{4}{\D}\delta\ta\left(r\D-r-1\right)\partial_a\partial^a\tr[\BT_\ta]\;,
\label{eq:deltaScal}
\end{equation}
for the scalar curvature. In \eqref{eq:deltaRic} and \eqref{eq:deltaScal} we used the additional constraint
\begin{equation}
    \partial^{a}\hT_{\ta,ab}=r\,\partial_{b}\tr[\BT_\ta]\;,
\end{equation}
coming from the conservation of the stress-energy tensor in flat space, i.e. $\partial_a\T_\ta^{ab}=0$. From \eqref{eq:deltaScal} it follows that
\begin{equation}
\label{eq:deltaScalnull}
    \delta R=0 \quad\Longleftrightarrow\quad r=\frac{1}{\D-1}\;.
\end{equation}
Let us consider separately the cases $\D=2$ and $\D>2$.
\begin{itemize}
    \item case $\D>2$: from \eqref{eq:deltaRiem}, it emerges that the deformation of the Riemann tensor depends on the field configuration through the stress-energy tensor and it is,in general, non-vanishing. Therefore, we conclude that the deformation induced by \eqref{eq:Od} modifies the geometry of the space in a non-trivial way for $\D>2$.
\item case $\D=2$: in this case the Riemann tensor has only one independent component, i.e. the scalar curvature $R$. From \eqref{eq:deltaScalnull} it follows that the operator $\cO_\ta^{[r,2]}$ modifies the geometry of the space for any $r\neq 1$. The case $r=1$ is special and corresponds to the  $\TbT$ operator $\cO_\ta^{\TbT}=\cO_\ta^{[1,2]}$ which does not affect the geometry, in agreement with the existence of a coordinate transformation.
\end{itemize}

\section{Metric flow equation}
In this section, we derive a system of differential equations that completely defines the flow of the metric. Moreover, we develop a perturbative algorithm to find a power series expansion for the solution to the metric flow equation.

The equivalence \eqref{eq:cAequivfin2} leads to the following system of differential equations,
 \begin{equation}
\label{eq:system}
\begin{cases}
    \dfrac{\dd g_{\mu\nu}}{\dd\var}=\dfrac{4}{\D}\hT_{\var,\mu\nu}\vspace{1mm}\\
    \dfrac{\partial \T_\var^{\mu\nu}}{\partial\var}=\dfrac{-2 }{\D\sqrt{g}}\dfrac{\partial}{\partial g_{\mu\nu}}\left(\sqrt{g}\,\hT_{\var,\rho\sigma}\T_\var^{\rho\sigma}\right)
\end{cases}\;,
\end{equation}
where the second equation descends from \eqref{eq:genericflow} and \eqref{eq:Hstressen}. Using the properties
\begin{equation}
    \dfrac{\partial g}{\partial g_{\mu\nu}} = g\,g^{\mu\nu}\;,
\end{equation}
and 
\begin{equation}
\dfrac{\partial\hT_{\var,\mu\nu}}{\partial g_{\rho\sigma}} = r\left(\delta_\mu^\rho\delta_\nu^\sigma \tr[\BT_\var]+ g_{\mu\nu}\T_\var^{\rho\sigma}\right)-\delta_\mu^\rho \T_{\var,\nu}^{\sigma}-\delta_\nu^\rho \T_{\var,\mu}^\sigma+f_{\mu\nu\alpha\beta}\dfrac{\partial\T_\var^{\alpha\beta}}{\partial g_{\rho\sigma}}\;,
\end{equation}
the second equation of \eqref{eq:system} yields explicitly
\begin{equation}
    \dfrac{\partial\T^{\mu\nu}_\var}{\partial \var}=\frac{4}{\D}\left[\T_\var^{2,\mu\nu}-r\,\T_\var^{\mu\nu}\tr[\BT_\var]-\dfrac14g^{\mu\nu}\left(r\,\tr[\BT_\var]^2-\tr[\BT^2_\var]\right)-\hT_{\var,\rho\sigma}\dfrac{\partial\T_\var^{\rho\sigma}}{\partial g_{\mu\nu}}\right]\;,
\end{equation}
where we denoted $\T_\var^{n,\mu\nu}=\T_\var^{\mu\mu_1}g_{\mu_1\mu_2}\T_\var^{\mu_2\mu_3}\dots g_{\mu_{n-1}\mu_n}\T_\var^{\mu_n\nu}$. The key point of the computation is that $\T_\var^{\mu\nu}$ depends on $\var$ both explicitly and implicitly through $g_{\mu\nu}$. Using the property
\begin{equation}
    \dfrac{\partial \T_\var^{\mu\nu}}{\partial g_{\rho\sigma}}=    \dfrac{1}{2}\left(g^{\mu\nu}\T_\var^{\rho\sigma}-g^{\rho\sigma}\T_\var^{\mu\nu}\right)+\dfrac{\partial \T_\var^{\rho\sigma} }{\partial g_{\mu\nu}}\;,
\end{equation}
we find that the total derivative of $\T_\var^{\mu\nu}$ w.r.t. $\var$ is
\begin{align}
    \dfrac{\dd\T_\var^{\mu\nu}}{\dd \var}&=\frac{\partial\T_\var^{\mu\nu}}{\partial \var}+\frac{\dd g_{\rho\sigma}}{\dd \var}\frac{\partial \T_\var^{\mu\nu}}{\partial g_{\rho\sigma}}\notag\\
&=\frac{4}{\D}\Biggl[\T_\var^{2,\mu\nu}-\frac12\left(dr+2r-1\right)\T_\var^{\mu\nu}\tr[\BT_\var]
    +\frac{g^{\mu\nu}}{4}\left(r\,\tr[\BT_\var]^2-\tr[\BT_\var^2]\right)\Biggr]\;.
\end{align}
From the latter expression, the first equation of \eqref{eq:system} and formula
\begin{equation}
\label{eq:dTrT}
    \frac{\dd\,\tr[\BT_\var]}{\dd\var}=\frac{\dd}{\dd\var}\left(g_{\mu\nu}\T_\var^{\mu\nu}\right)=\left(\frac{2}{\D}-r\right)\tr[\BT_\var]^2-\tr[\BT_\var^2]\;,
\end{equation}
we can easily compute the total derivative of $\hT_{\var,\mu\nu}$ as
\begin{equation}
    \dfrac{\dd\hT_{\var,\mu\nu}}{\dd \var} = \dfrac{\dd}{\dd \var}\left(r\,g_{\mu\nu}\tr[\BT_\var]-g_{\mu\rho}\T_\var^{\rho\sigma}g_{\nu\sigma}\right)\;,
\end{equation}
Upon explicit computation, we arrive to the system
\begin{equation}
\label{eq:systemfin}
\begin{cases}
    \dfrac{\dd g_{\mu\nu}}{\dd\var}=\dfrac{4}{\D}\hT_{\var,\mu\nu}\vspace{1mm}\\
    \dfrac{\dd\hT_{\var,\mu\nu}}{\dd \var}=\dfrac{4}{\D}\hT_{\var,\mu\nu}^2+\alpha_\var\hT_{\var,\mu\nu}+\beta_\var g_{\mu\nu}
    \end{cases}\;,
\end{equation}
where we denoted $\hT_{\var,\mu\nu}^n=\hT_{\var,\mu\mu_1}g^{\mu_1\mu_2}\hT_{\var,\mu_2\mu_3}\dots g^{\mu_{n-1}\mu_{n}}\hT_{\var,\mu_n\nu}$ and we defined
\begin{equation}
\label{eq:alphabetadef}
\alpha_\var=\frac{2}{\D}\left(1-\D r\right)\tr[\BT_\var]\;,\quad\beta_\var=\frac{\D r-1}{\D}\left(r\,\tr[\BT_\var]^2-\tr[\BT_\var^2]\right)\;.
\end{equation}
The system \eqref{eq:systemfin} completely defines the flow of the metric once an initial condition has been chosen. The idea is to solve it for $g_{\mu\nu}(\var):=g_{\mu\nu}(\var;\var_0)$ with initial condition $g_{\mu\nu}(\var_0)=\eta_{ab}\,e_\mu^a\,e_\nu^b$ for some $\var_0$, where $e^a_\mu=\delta^a_\mu$ is the trivial vierbein. Such solution provides the deformed background metrics that allow to boost or absorb the deformation of the action, depending on the choice of the parameters $\var_0$ and $\var$. In fact, 
\begin{enumerate}
    \item[1)] if $\var_0=\ta$ and $\var=\ta_0$: the action $\cA_\ta$ with metric $\eta_{ab}$ is dynamically equivalent to $\cA_{\ta_0}$ with metric 
    \begin{equation}
    \label{eq:hg}
        \hg_{\mu\nu}(\ta)=g_{\mu\nu}(\ta_0;\ta)\;;
    \end{equation}
    \item[2)] if $\var_0=\ta_0$ and $\var=\ta$: the action $\cA_{\ta_0}$ with metric $\eta_{ab}$ is dynamically equivalent to $\cA_\ta$ with metric 
    \begin{equation}
    \label{eq:cg}
        \cg_{\mu\nu}(\ta)=g_{\mu\nu}(\ta;\ta_0)\;.
    \end{equation}
\end{enumerate} 

\subsection{Algorithm to solve the metric flow equation}
\label{sec:metricflow}
In this section we compute the solution $g_{\mu\nu}(\var):=g_{\mu\nu}(\var;\var_0)$ of \eqref{eq:systemfin} by means of a perturbative approach which can be made algorithmic and implemented in a computer software.

The idea is to Taylor expand $g_{\mu\nu}(\var):=g_{\mu\nu}(\var;\var_0)$ around $\var=\var_0$ as 
\begin{equation}
\label{eq:Taylor}
    g_{\mu\nu}(\var)=\sum_{n=0}^\infty \frac{g_{ab}^{(n)}(\var_0)}{n!}\left(\var-\var_0\right)^n e^a_\mu\, e^b_\nu \;,
\end{equation}
where $e^a_\mu=\delta^a_\mu$ is the trivial vierbein and
\begin{equation}
    g_{\mu\nu}^{(n)}(\var)=\frac{\dd^n g_{\mu\nu} }{\dd \var^n}\;,\quad g_{ab}^{(n)}(\var_0)=e^\mu_a\, e^\nu_b\left.\frac{\dd^n g_{\mu\nu} }{\dd \var^n}\right|_{\var=\var_0}\;.
\end{equation}
We impose $g_{ab}^{(0)}(\var_0)=\eta_{ab}$ and look for the coefficients $\{g_{ab}^{(n)}(\var_0)\}_{n\geq 1}$. The first two coefficients $g_{ab}^{(1)}(\var_0)$ and $g_{ab}^{(2)}(\var_0)$ descend trivially from \eqref{eq:system} and yields
\begin{align}
\label{eq:gord1}
    &g_{ab}^{(1)}(\var_0)=\frac{4}{\D}\hT_{\var_0,ab}\;,\\
    \label{eq:gord2}
    &g_{ab}^{(2)}(\var_0)=\left(\frac{4}{\D}\right)^2\hT_{\var_0,ab}^2+\frac{4\alpha_{\var_0}}{\D}\hT_{\var_0,ab}+\frac{4\beta_{\var_0}}{\D}\eta_{ab}\;,
\end{align}
where $\alpha_\var$ and $\beta_\var$ are defined as per \eqref{eq:alphabetadef}. To get $\{g_{ab}^{(n)}(\var_0)\}_{n\geq 3}$ we need to find a strategy to compute $g_{\mu\nu}^{(n)}$ from $g_{\mu\nu}^{(n-1)}$. Using \eqref{eq:systemfin} and the flow equation for the inverse metric 
\begin{equation}
    \frac{\dd g^{\mu\nu}}{\dd\var}=-\frac{4}{\D}\hT_\var^{\mu\nu}\;,
\end{equation} 
we arrive at the recurrence relation
\begin{equation}
    \frac{\dd\hT_{\var,\mu\nu}^k}{\dd\var}=\frac{\dd}{\dd\var}\left(\hT_{\var,\mu\rho}g^{\rho\sigma}\hT_{\var,\sigma\nu}^{k-1}\right)
    =\alpha_\var\hT_{\var,\mu\nu}^k+\beta_\var\hT_{\var,\mu\nu}^{k-1}+\hT_{\var,\mu\rho}g^{\rho\sigma}\frac{\dd\hT_{\var,\sigma\nu}^{k-1}}{\dd\var}\;,\quad\forall k\geq 2\;,
\end{equation}
which gives
\begin{equation}
\label{eq:hTk}
    \frac{\dd\hT_{\var,\mu\nu}^k}{\dd\var}=\frac{4}{\D}\hT_{\var,\mu\nu}^{k+1}+k\alpha_\var\hT_{\var,\mu\nu}^k+k\beta_\var\hT_{\var,\mu\nu}^{k-1}\;,\quad\forall k\geq 2\;.
\end{equation}
Using \eqref{eq:dTrT}, formula
\begin{equation}
\label{eq:dTrT2}
    \frac{\dd\,\tr[\BT_\var^2]}{\dd\var}=\frac{\dd}{\dd\var}\left(g_{\mu\nu}\T_\var^{\mu\rho}g_{\rho\sigma}\T_\var^{\sigma\nu}\right)=\frac{2}{\D}\tr[\BT_\var] \left(r\, \tr[\BT_\var]^2+\left(1-2\D r\right) \tr[\BT_\var^2]\right)\;,
\end{equation}
and the definition \eqref{eq:alphabetadef}, we arrive at\footnote{Notice that \eqref{eq:dalphabeta} is a system of differential equations that can be exactly solved for $\alpha_\var$ and $\beta_\var$ as functions of $\var$. Currently, it is unclear to us whether such explicit solution might be helpful in computing the metric or other quantities such as the action.}
\begin{equation}
\label{eq:dalphabeta}
   \frac{\dd\alpha_\var}{\dd\var}=
   \alpha_\var^2-2\beta_\var\;,\quad \frac{\dd\beta_\var}{\dd\var}=
   \alpha_\var\beta_\var\;.
\end{equation}
Formulae \eqref{eq:hTk} and \eqref{eq:dalphabeta} imply that $g_{\mu\nu}^{(n)}$ can be written as
\begin{equation}
\label{eq:gexp}
    g_{\mu\nu}^{(n)}=c_0^{(n)} g_{\mu\nu}+\sum_{k=1}^n c_k^{(n)}\hT_{\var,\mu\nu}^k\;,\quad \forall n\geq 1\;,
\end{equation}
where $\{c_k^{(n)}\}_{k\in\{1,\dots,n\}}$ are polynomials in the variables $\alpha_\var$ and $\beta_\var$ with real coefficients. Therefore, the computation of $g_{\mu\nu}^{(n)}$ has been reduced to the computation of the coefficients $\{c_k^{(n)}\}_{k\in\{1,\dots,n\}}$. Differentiating \eqref{eq:gexp} w.r.t. $\var$ and using \eqref{eq:hTk}, we easily obtain the recurrence relations
\begin{equation}
\label{eq:coeffrecursive}
    \begin{cases}
    \displaystyle{c_0^{(n+1)}=\frac{\dd c_0^{(n)}}{\dd\var}+\beta_\var c_1^{(n)}}\vspace{1mm}\\
    \displaystyle{c_{k}^{(n+1)}=\frac{4}{\D}c_{k-1}^{(n)}+\frac{\dd c_{k}^{(n)}}{\dd\var}+k\alpha_\var c_{k}^{(n)}+\left(k+1\right)\beta_\var c_{k+1}^{(n)} \;,\quad 1\leq k\leq n-1}\vspace{1mm}\\
    \displaystyle{c_n^{(n+1)}=\frac{4}{\D}c_{n-1}^{(n)}+\frac{\dd c_{n}^{(n)}}{\dd\var}+n\alpha_\var c_{n}^{(n)}}\vspace{1mm}\\
    \displaystyle{c_{n+1}^{(n+1)}=\frac{4}{\D}c_{n}^{(n)}}
    \end{cases}\;,
\end{equation}
where
\begin{equation}
    \frac{\dd c_k^{(n)}}{\dd\var}=\frac{\partial c_k^{(n)}}{\partial\alpha_\var}\frac{\dd\alpha_\var}{\dd\var}+\frac{\partial c_k^{(n)}}{\partial\beta_\var}\frac{\dd\beta_\var}{\dd\var}\;.
\end{equation}
Formula \eqref{eq:coeffrecursive} allows to recover (the coefficients of) $\{g_{\mu\nu}^{(n)}\}_{n\geq 3}$ from the initial condition at $n=2$
\begin{equation}
    c_0^{(2)}=\left(\frac{4}{\D}\right)^2\;,\quad c_1^{(2)}=\frac{4\alpha_\var}{\D}\;,\quad c_2^{(2)}=\frac{4\beta_\var}{\D}\;.
\end{equation}
For example, the coefficients of $g_{\mu\nu}^{(3)}$ are
\begin{equation}
    c_0^{(3)}=\frac{8\alpha_\var\beta_\var}{d}\;,\quad c_1^{(3)}=\frac{8}{d^2} \left(\alpha_\var^2 d-\beta_\var (d-6)\right)\;,\quad
    c_2^{(3)}=\frac{48\alpha_\var}{d^2}\;,\quad c_3^{(3)}=\left(\frac{4}{\D}\right)^3\;.
\end{equation}
The implementation of the recurrence relations \eqref{eq:coeffrecursive} corresponds to a couple of lines in a Mathematica notebook. The first $n=100$ terms of the sequence $\{c_k^{(n)}\}_{k\in\{1,\dots,n\}}$ can be obtained in less than a minute on a standard laptop. However, the task of finding a close expression for $g_{\mu\nu}^{(n)}$ valid for all $n\in\mathbbm{N}$ is highly non-trivial. In the next section we shall consider special cases in which this task becomes feasible.

\subsection{Exact solutions for the metric}
\label{sec:metrictrunc}
In this section we show that for some values of $r$ and under some assumptions on the stress-energy tensor, it is possible to obtain a close expression for the coefficient $g_{ab}^{(n)}(\var_0)$ valid for all $n\geq 1$ and we are able to formally sum the series \eqref{eq:Taylor}.

It is convenient to work with the matrix notation. Let us introduce the $\D\times \D$ matrices
\begin{equation}
    \BG^{(n)}(\var)=\left(g^{\mu\rho}g_{\rho\nu}^{(n)}\right)_{\mu,\nu\in\{0,\dots,\D-1\}}\;,\quad n\geq 1\;.
\end{equation}
Assume that the matrix $\BT_{\var_0}$ is diagonalisable, i.e. there exist an invertible matrix $\BP$ and a diagonal matrix $\BD$ such that $\BT_{\var_0}=\BP\BD\BP^{-1}$. Moreover, assume that $\BT_{\var_0}$ has 2 (resp. 1) independent eigenvalues of multiplicity $\frac{\D}{2}$ (resp. $\D$) if $\D$ is even (resp. odd), namely
\begin{equation}
\label{eq:Ddiag}
\BD=
\begin{cases}
\text{diag}\,\bigl(\underbrace{\lambda_1,\dots,\lambda_1}_\text{$\frac{\D}{2}$-times},\underbrace{\lambda_2,\dots,\lambda_2}_\text{$\frac{\D}{2}$-times}\bigr) \;,\quad \D\in2\mathbb{N}+2 \\
\text{diag}\,\bigl(\underbrace{\lambda,\dots,\lambda}_\text{$\D$-times}\bigr) \;,\quad \D\in2\mathbb{N}+3
\end{cases}\;.
\end{equation}
\subsubsection{Case $r=\frac{2}{\D}$}
Under the assumption \eqref{eq:Ddiag} one can show that, for all $n\geq 1$
\begin{equation}
\BP^{-1}\BG^{(n)}(\var_0)\,\BP=(-1)^n\left(-\frac{4}{\D}\right)_n\left(\frac{2}{\D}\tr[\BD]\Id_{\D}-\BD\right)^n\;,
\end{equation}
where $(x)_n=\frac{\Gamma(x+n)}{\Gamma(x)}$ is the Pochhammer symbol. Whence
\begin{equation}
\label{eq:coeffgn}
g_{ab}^{(n)}(\var_0)=(-1)^n\left(-\frac{4}{\D}\right)_n\hT_{\var_0,ab}^n\;,
\end{equation}
and \eqref{eq:Taylor} can be formally written as
\begin{equation}
\label{eq:gsol}
g_{\mu\nu}(\var;\var_0)=\left[\left(\eta+\left(\var-\var_0\right)\hT_{\var_0}\right)^{\frac{4}{\D}}\right]_{ab}e^a_\mu\, e^b_\nu\;.
\end{equation}
Observe that, differentiating both sides of \eqref{eq:gsol} w.r.t. $\var$ and using the first equation of \eqref{eq:system}, we find
\begin{equation}
\label{eq:hTsol}
    \hT_{\var,\mu\nu} = \hT_{\var_0,ac}\left[\left(\eta+\left(\var-\var_0\right)\hT_{\var_0}\right)^{\frac{4}{\D}-1}\right]_{ib} \eta^{ci}\, e^a_\mu \,e^b_\nu\;.
\end{equation}
Let us make a few remarks:
\begin{itemize}
    \item in $\D=2$, the condition \eqref{eq:Ddiag} does not constraint the stress-energy tensor which has, in general, 2 distinct eigenvalues. Therefore, \eqref{eq:gsol} is the deformed metric associated to the $\TbT$ deformation of a generic theory in $\D=2$. Moreover, using the identifications 
    \begin{equation}
        \T_{\var_0,0}^0=-\mathcal{H}_{\var_0}\;,\quad \T_{\var_0,0}^1=\T_{\var_0,1}^0=\mathbbm{i}\mathcal{P}_{\var_0}\;,
    \end{equation}
    where $\mathcal{H}_{\var_0}$ and $\mathcal{P}_{\var_0}$ are the energy and momentum densities and setting $\dd x^0=0$, the line element $\dd\ell^2=g_{\mu\nu}(\var;\var_0)\,\dd x^\mu \dd x^\nu$ becomes
    \begin{equation}
    \label{eq:lineel2D}
       \dd\ell^2=\left[\left(1-\left(\var-\var_0\right)\mathcal{H}_{\var_0}\right)^2-\left(\left(\var-\var_0\right) \mathcal{P}_{\var_0}\right)^2\right]\bigl(\dd x^1\bigr)^2\;.
    \end{equation}
    Formula \eqref{eq:lineel2D} resembles the modification of the ``effective size'' of the system at quantum level (see, for example, equation (2.8) in \cite{Conti:2018tca}), which is ultimately a consequence of the Zamolodchikov's factorisation Theorem \cite{Zamolodchikov:2004ce}.
    \item in $\D=4$, \eqref{eq:gsol} takes a particularly simple expression, being linear in $\var$. It is natural to ask whether also in this case the information of the quantum theory is hidden in the line element $\dd\ell^2$, in analogy to the $\D=2$ case.
    \item all formulas can be analytically continued to $\D=1$, in which the field theory reduces to a mechanical system. In this case, the tensors $g_{\mu\nu}$ and $\T_{\var,\nu}^\mu$ reduce to the scalars $g_{00}=g$ and $T_{\var,0}^0:=-E_\var$ respectively, where $E_\var$ is the energy, while the perturbing operator is $\cO_\ta^{[2,1]}=E_\ta^2$ (see \cite[Appendix A]{Gross:2019ach}). Moreover, from \eqref{eq:hTsol} using $\hT_{\var,00}=-gE_\var$ we get
    \begin{equation}
        E_\var=\frac{E_{\var_0}}{1-(\var-\var_0)E_{\var_0}}\;,
    \end{equation}
    which matches the result of \cite{Gross:2019ach}. Notice that this is also the expression of the deformed energy density of a Yang-Mills theory in $\D=2$ \cite{Conti:2018jho,Santilli:2020qvd,Griguolo:2022xcj}.
    \item it would be interesting to look for a match between the series expansion of the metric proposed here and the perturbative results obtained in \cite{Babaei-Aghbolagh:2020kjg} for the Lagrangians associated to abelian gauge theories in $d\in2\mathbbm{N}$ deformed by the operator $\cO_\ta^{[\frac{2}{d},d]}$.\footnote{We thank Hossein Babaei-Aghbolagh for suggesting to us the possibility to perform this comparison.}
\end{itemize}

\subsubsection{Case $r=\frac{1}{\D}$}
Under the assumption \eqref{eq:Ddiag} one can show that, for all $n\geq 1$
\begin{equation}
\BP^{-1}\BG^{(n)}(\var_0)\,\BP=\left(\frac{4}{\D}\right)^n\left(\frac{1}{\D}\tr[\BD]\Id_{\D}-\BD\right)^n\;.
\end{equation}
Notice that the RHS of the latter equation is identically zero if $d$ is odd, due to the definition \eqref{eq:Ddiag}. Then, restricting to the non trivial case in which $d$ is even we have
\begin{equation}
g_{ab}^{(n)}(\var_0)=
\left(\frac{4}{\D}\right)^n\hT_{\var_0,ab}^n \;,\quad d\in2\mathbbm{N}+2\;,
\end{equation}
and \eqref{eq:Taylor} can be formally written as
\begin{equation}
\label{eq:gsol2}
g_{\mu\nu}(\var;\var_0)=
\left[\exp\left(\frac{4}{\D}\left(\var-\var_0\right)\hT_{\var_0}\right)\right]_{ab}e^a_\mu\, e^b_\nu \;,\quad d\in2\mathbbm{N}+2\;.
\end{equation}
Let us stress again that, since the condition \eqref{eq:Ddiag} does not constraint the stress-energy tensor in $d=2$, the latter solution holds for any theory in $d=2$ deformed by the operator $\cO_\ta^{[\frac12,2]}$.\\

The existence of an exact solution for the metric suggests that the action deformed by the operator $\cO_\ta^{[\frac{1}{\D},\D]}$ might have a close expression as well. In the following, we will address the computation of the deformed Lagrangian density focusing on the simple case of a non-interacting scalar field in $\D=2$. We leave the analysis of more complicated theories in $\D=2$ as well as the extension to theories in $\D>2$ to a future publication.

It is possible to show that the solution to the flow equation 
\begin{equation}
    \frac{\partial\cL_\ta}{\partial\ta}=\cO_\ta^{[\frac12,2]}\;,
\end{equation}
with initial condition
\begin{equation}
\cL_0 = \frac12  \eta^{ab}\partial_a\phi\,\partial_b\phi\;,
\end{equation}
is given by
\begin{equation}
\label{eq:W}
    \cL_\ta=\frac{W(2\ta\cL_0)}{2\ta}\left( \frac12 W(2\ta\cL_0) +1\right)\;,
\end{equation}
where $W(x)$ is the Lambert function. It is surprising that 
\begin{equation}
F(x)=  - W(-x) \left( \frac12 W(-x) +1 \right) = \sum_{n=1}^{\infty} \frac{n^{n-2}}{n!} x^n\;,
\end{equation}
corresponds to the generating function associated to Cayley’s formula in graph theory.\footnote{Cayley's formula 
states that there are $n^{n-2}$ labeled trees on $n$ vertices (see, for example \cite{mezo2022lambert}).}

\subsection{Abelian gauge theories in $\D=4$: the exact vierbein}
\label{sec:abelian}
The assumption \eqref{eq:Ddiag} restricts the range of applicability of the results obtained in section \ref{sec:metrictrunc} for $\D>2$. However, in $\D=4$ there exists a whole class of field theories whose stress-energy tensors fulfil the constraint \eqref{eq:Ddiag}: the abelian gauge theories, describing the dynamics of one gauge field, i.e. the electromagnetic four-potential $A_a$ (in flat space with metric $\eta_{ab}$).

We briefly review the proof of this fact, which can be found also in \cite[Appendix A]{Ferko:2022iru}. Let $F_{ab}=\partial_a A_b-\partial_b A_a$ be the field-strength associated to the gauge field and $\tF_{ab}=\frac12\epsilon_{abij}F^{ij}$ the dual field-strength, where $\epsilon_{abij}$ is the Levi-Civita symbol with the choice $\epsilon_{0123}=1$. Following \cite{Ferko:2022iru}, the stress-energy tensor of a generic abelian gauge theory can be decomposed as
\begin{equation}
\label{eq:Tabelian}
    \T_{ab}=a^{(0)}\eta_{ab}+a^{(1)}F_{ab}^2 +a^{(2)}F_{ab}^4\;,
\end{equation}
where $a^{(0)}$, $a^{(1)}$ and $a^{(2)}$ are functions of $\tr[\BF^2]$ and $\tr[\BF^4]$ with $\BF=\bigl(F^a_b\bigr)_{a,b\in\{0,\dots,3\}}$.

A straightforward computation shows that the eigenvalues $\{\lambda_i\}_{i\in\{1,\dots,4\}}$ of the matrix $\BF$ are such that $\lambda_2=-\lambda_1$ and $\lambda_4=-\lambda_3$ (independently of the signature of the metric $\eta_{ab}$) whence it follows that the eigenvalues $\{\bar{\lambda}_i\}_{i\in\{1,\dots,4\}}$ of $\BT=\{\T^a_b\}_{a,b\in\{0,\dots,3\}}$ are \begin{equation}
    \bar\lambda_i=a^{(0)}+a^{(1)}\lambda_i^2+a^{(2)}\lambda_i^4\;,
\end{equation}
with $\bar\lambda_2=\bar\lambda_1$ and $\bar\lambda_4=\bar\lambda_3$.

We conclude that, for any representative of the family of abelian gauge theories belonging to the flow of $\cO_\ta^{[\frac12,4]}$ -- a notable example being ModMax Born-Infeld (see section \ref{sec:MMBIdimred}) -- the deformed metrics \eqref{eq:hg} and \eqref{eq:cg} are given by
\begin{align}
    &\hg_{\mu\nu}(\ta)=\left(\eta_{ab}-\left(\ta-\ta_0\right)\hT_{\ta,ab}\right)e_\mu^a e_\nu^b\;,\notag\\
    &\cg_{\mu\nu}(\ta)=\left(\eta_{ab}+\left(\ta-\ta_0\right)\hT_{\ta_0,ab}\right)e_\mu^a e_\nu^b\;,
    \label{eq:hgcgAbelian}
\end{align}
where the tensors $\hT_{\ta,ab}$ and $\hT_{\ta_0,ab}$ are both evaluated in flat space with metric $\eta_{ab}$. Notice that the expression of \eqref{eq:hgcgAbelian} matches that of the (pseudo) metric found many years ago (see the lecture notes \cite{Plebanski:1970zz}) for the Maxwell Born-Infeld theory, using a completely different approach.  

From the discussion of section \ref{sec:defcurv} we know that, in general, \eqref{eq:hgcgAbelian} are curved. In appendix \ref{sec:appendix}, we show that it is possible to give an exact expression for the pair of vierbein $\hat{e}_\mu^a:=\hat{e}_\mu^a(\ta)$ and $\check{e}_\mu^a:=\check{e}_\mu^a(\ta)$ that fulfil
\begin{equation}
\label{eq:hgcgjac}
    \hg_{\mu\nu}(\ta)=\hat{e}^a_\mu \,\hat{e}^b_\nu\, \eta_{ab} \;,\quad \cg_{\mu\nu}(\ta)=\check{e}^a_\mu\, \check{e}^b_\nu\, \eta_{ab}\;.
\end{equation}
Decomposing $\T_{\ta_0,ab}$ and $\T_{\ta,ab}$ as per \eqref{eq:Tabelian} with coefficients $\left(a_{\ta_0}^{(0)},a_{\ta_0}^{(1)},a_{\ta_0}^{(2)}\right)$ and $\left(a_{\ta}^{(0)},a_{\ta}^{(1)},a_{\ta}^{(2)}\right)$, respectively, we obtain
\begin{align}
    \hat{e}_\mu^a(\ta)&=\sqrt{\Sigma\left(\ta_0;\ta\right)}\,e_\mu^a+\sqrt{\ta_0-\ta}\left(u^{(1)}_\ta F_{cb}+u^{(2)}_\ta F^3_{cb}\right)\eta^{ac} e^b_\mu\;,\notag\\
    \check{e}_\mu^a(\ta)&=\sqrt{\Sigma\left(\ta;\ta_0\right)}\,e_\mu^a+\sqrt{\ta-\ta_0}\left(u^{(1)}_{\ta_0}F_{cb}+u^{(2)}_{\ta_0}F^{3}_{cb}\right)\eta^{ac}e^b_\mu\;,\label{eq:hcJ}
\end{align}
where we defined
\begin{align}
    &u^{(1)}_{\var}=\frac{\sqrt{2}\left(a^{(1)}_{\var}+\frac{\cV_{\var}}{4}\right)}{\sqrt{4a^{(1)}_{\var}+\cV_{\var}+a^{(2)}_{\var}\tr[\BF^2]}} \;,\quad u^{(2)}_{\var}=\frac{\sqrt{2} a^{(2)}_{\var}}{\sqrt{4a^{(1)}_{\var}+\cV_{\var}+a^{(2)}_{\var}\tr[\BF^2]}}\;,\notag\\
    &\Sigma(\var;\var_0)=1+\left(\var-\var_0\right)\left(a^{(0)}_{\var_0}+\frac12 a^{(1)}_{\var_0}\tr[\BF^2]+\frac12 a^{(2)}_{\var_0}\tr[\BF^4]\right)\;,
    \label{eq:urelations}
\end{align}
and
\begin{equation}
    \cV_{\var}=\sqrt{\bigl(4a^{(1)}_{\var}\bigr)^2+8 a^{(1)}_{\var} a^{(2)}_{\var} \tr[\BF^2]+2\bigl(a^{(2)}_{\var}\bigr)^2\left(\tr[\BF^2]^2-2 \tr[\BF^4]\right)}\;.
\end{equation}
We observe that if $a^{(2)}_{\var}= 0$ relations \eqref{eq:urelations} drastically simplify and \eqref{eq:hcJ} reduces to
\begin{align}
    &\hat{e}_\mu^a(\ta)=
    \sqrt{\sigma(\ta_0;\ta)}\,e_\mu^a+\sqrt{\left(\ta_0-\ta\right)a^{(1)}_\ta}\,F^a_b e_\mu^b\;,\notag\\
    &\check{e}_\mu^a(\ta)=\sqrt{\sigma(\ta;\ta_0)}\,e_\mu^a+\sqrt{\left(\ta-\ta_0\right)a^{(1)}_{\ta_0}}\,F^a_b e_\mu^b\;,
    \label{eq:hcJMM}
\end{align}
with
\begin{equation}
   \sigma(\var;\var_0)=1+\left(\var-\var_0\right)\left(a^{(0)}_{\var_0}+\frac{1}{2}a^{(1)}_{\var_0}\,\tr[\BF^2]\right)\;.
\end{equation}
Relevant examples of models such that $a^{(2)}_{\var}=0$ are ModMax and its Born-Infeld-like extension, that we shall briefly discuss in the next section.

\section{ModMax and its Born-Infeld-like extension}
\label{sec:MMBIdimred}
Let us recall that the ModMax (MM) theory \cite{Bandos:2020jsw} represents a marginal deformation of the Maxwell theory described by the Euclidean action $\cA_\gamma^{\MM}=\int \dd^4\Bx\,\cL_\gamma^{\MM}$ with
\begin{align}
\label{eq:MMLag}
\cL_\gamma^{\MM}=\cosh(\gamma)\,S- \sinh(\gamma)\sqrt{S^2-P^2} \;,
\end{align}
where $\gamma$ is a real parameter and we defined the invariants\footnote{Notice that the definition of the invariants \eqref{eq:SPdef} becomes in Minkowsky signature $S_M=-S$ and $P_M:=-\mathbbm{i} P$. Correspondingly, the Lagrangian density in Minkowsky signature becomes $\cL_M^{\MM}(\gamma)=-\cL^{\MM}(\gamma)$.}
\begin{align}
\label{eq:SPdef}
S:=\frac14 F_{ab}F^{ab} \;,\quad P:=\frac14\tF_{ab}F^{ab}=\sqrt{\det[\BF]}\;.
\end{align}
One can associate to \eqref{eq:MMLag} a Born-Infeld-like extension (MMBI) \cite{Bandos:2020hgy} that is described by the action $\cA_{\ta,\gamma}^{\MMBI}=\int \dd^4\Bx\,\cL_{\ta,\gamma}^{\MMBI}$ with
\begin{align}
\label{eq:MMBILag}
\cL_{\ta,\gamma}^{\MMBI}=\frac{-1+\cS^{\MMBI}}{2\ta}\;,\quad \cS^{\MMBI}=\sqrt{1+4\ta\cL_\gamma^{\MM}+4\ta^2P^2}\;.
\end{align}
Clearly $\cA_0^{\MM}=\cA^{\M}$ and $\cA_{\ta,0}^{\MMBI}=\cA_\ta^{\MBI}$ where $\cA^{\M}$ and $\cA_\ta^{\MBI}$ are the Maxwell (M) and the Maxwell Born-Infeld (MBI) action respectively. A simple computation shows that the components of the stress-energy tensor associated to \eqref{eq:MMBILag} take the simple form
\begin{equation}
\label{eq:TMMBI}
    (\T_{\ta,\gamma}^{\MMBI})_{ab}=a^{\MMBI}_{\ta,\gamma} \eta_{ab}+b^{\MMBI}_{\ta,\gamma} F_{ab}^2 \;,
\end{equation}
with coefficients
\begin{align}
    &a^{\MMBI}_{\ta,\gamma}=\frac{P^2}{\cS^{\MMBI}}\left(2\ta+\frac{\sinh (\gamma )}{\sqrt{S^2-P^2}}\right)+\frac{1-\cS^{\MMBI}}{2 \ta } \;,\notag \\ &b^{\MMBI}_{\ta,\gamma}=\frac{1}{\cS^{\MMBI}}\left(\frac{S \sinh (\gamma )}{\sqrt{S^2-P^2}}-\cosh (\gamma )\right)\;.
    \label{eq:TMMBIcoeff}
\end{align}
Using \eqref{eq:TMMBI} and \eqref{eq:TMMBIcoeff} it is possible to show (see \cite{Ferko:2022iru}) that the action $\cA_{\ta,\gamma}^{\MMBI}$ belongs to the flow of the irrelevant operator $\cO_\ta^{[\frac12,4]}$ with initial condition $\cA_\gamma^{\MM}$ at $\ta=0$ for any value of $\gamma$.

Interestingly, in \cite{Babaei-Aghbolagh:2022uij} it was made the important observation that the action $\cA_{\ta,\gamma}^{\MMBI}$ belongs to the flow of the marginal operator
\begin{equation}
    \widetilde{\cO}_\gamma^{[4]}:=\sqrt{-\cO_\gamma^{[\frac14,4]}}=\frac12\sqrt{\tr[\BT_\gamma^2]-\frac14 \tr[\BT_\gamma]^2}\;,
\end{equation}
with initial condition $\cA_{\ta}^{\MBI}$ at $\gamma=0$ for any value of $\ta$. This fact has been shown in \cite{Babaei-Aghbolagh:2022uij} by means of a perturbative expansion around $\gamma=0$. Let us briefly report here the exact computation at finite values of $\gamma$. Using \eqref{eq:TMMBI} and \eqref{eq:TMMBIcoeff} one has
\begin{equation}
    \frac12\sqrt{\tr[(\BT_{\ta,\gamma}^{\MMBI})^2]-\frac14\tr[\BT_{\ta,\gamma}^{\MMBI}]^2} = b_{\ta,\gamma}^{\MMBI}\sqrt{S^2-P^2} \;.
\end{equation}
On the other hand,
\begin{equation}
    \frac{\partial\cL_{\ta,\gamma}^{\MMBI}}{\partial\gamma} = \frac{1}{\cS^{\MMBI}}\left(S\sinh(\gamma)-\cosh(\gamma)\sqrt{S^2-P^2}\right)=b_{\ta,\gamma}^{\MMBI}\sqrt{S^2-P^2}\;,
\end{equation}
whence the equivalence
\begin{equation}
   \frac{\partial\cL_{\ta,\gamma}^{\MMBI}}{\partial\gamma}= \frac12\sqrt{\tr[(\BT_{\ta,\gamma}^{\MMBI})^2]-\frac14\tr[\BT_{\ta,\gamma}^{\MMBI}]^2} \;.
\end{equation}

\subsection{Dimensional reduction from $\D=4$ to $\D=2$}
It is well known \cite{Barbashov:1967zzz} that there is a deep connection between the theories of Nambu-Goto in $\D=2$ and Maxwell Born-Infeld in $\D=4$. In fact, particular solutions of Maxwell Born-Infeld are also solutions of Nambu-Goto in static gauge with two transversal scalar fields. In this section, we show that this link can be lifted to the Born-Infeld-like extension of ModMax. 

In analogy with \cite{Barbashov:1967zzz}, we consider a specific field configuration consisting in the scattering of plane waves along the direction $x^1$, which corresponds to the requirements
\begin{equation}
\label{eq:planewavecond}
A_\mu:=A_\mu(\bar\Bx)\;,\quad \partial_1 A_0 -\partial_0 A_1 = 0\;,
\end{equation}
where $\bar\Bx=(x^0,x^1)$ denotes the restricted set of local coordinates on the plane. Let us identify $\phi_a(\bar\Bx):=A_{a+1}(\bar\Bx)$ with $a\in\{1,2\}$. Then, the constraint \eqref{eq:planewavecond} implies the following reduction $F_{ab}(\Bx)\to \bar{F}_{ab}(\bar\Bx)$, where $\bar{F}_{ab}$ has only four non-vanishing (independent) components that depends on the derivative of the scalar fields $\{\phi_i\}_{i\in\{1,2\}}$ w.r.t. $\bar\Bx$: \begin{equation}
    \bar{F}_{02}=\partial_0\phi_1\;,\quad \bar{F}_{03}=\partial_0\phi_2\;,\quad \bar{F}_{12}=\partial_1\phi_1\;,\quad \bar{F}_{13}=\partial_1\phi_2\;.
\end{equation}
Consequently, the invariants $(S,P)$ as per \eqref{eq:SPdef} reduce to $(\bS_2,\bP_2)$, where we defined
\begin{equation}
\bS_N:=\cL^{{\Sc},N} \;,\quad \bP_N:=-\sqrt{\det[\BH_N]}\;,
\end{equation} 
with $H_{N,ab}$ the following symmetric tensor 
\begin{equation}
    H_{N,ab}=\sum_{i=1}^N\partial_a\phi_i\,\partial_b\phi_i\;,\quad \BH_N=\bigl(H_{N,b}^a\bigr)_{a,b\in\{0,1\}}\;,
\end{equation}
and 
\begin{equation}
    \cL^{{\Sc},N}=\frac12\tr[\BH_N]=\frac12\sum_{i=1}^N \eta^{ab}\partial_a\phi_i\,\partial_b\phi_i\;,
\end{equation}
the Lagrangian density describing $N$ non-interacting and massless scalar fields $\{\phi_i\}_{i\in\{1,\dots,N\}}$ in $\D=2$. Performing the transformation $(S,P)\to(\bS_2,\bP_2)$ in \eqref{eq:MMLag} and \eqref{eq:MMBILag} we obtain
\begin{equation}
   \left(\cL_\gamma^{\MM},\cL_{\ta,\gamma}^{\MMBI}\right)\to\left(\cL_\gamma^{{\MS},2},\cL_{\ta,\gamma}^{{\MSBI},2}\right) \;,
\end{equation}
where we defined the following family of Modified Scalar (MS) theory involving $N\geq 1$ scalar fields $\{\phi_i\}_{i\in\{1,\dots,N\}}$ with Lagrangian density\footnote{Notice that for $N=1$ one obtains $\cL_\gamma^{{\MS},1}=e^{-\gamma}\cL^{{\Sc},1}$.}
\begin{equation}
\label{eq:MSNLag}
    \cL_\gamma^{{\MS},N}=\cosh(\gamma)\,\bS_N-\sinh(\gamma)\sqrt{\bS_N^2-\bP_N^2}\;,
    \end{equation}
    and its Born-Infeld-like extension (MSBI)
    \begin{equation}
    \cL_{\ta,\gamma}^{{\MSBI},N}=\frac{-1+\cS^{{\MSBI},N}}{2\ta}\;,\quad \cS^{{\MSBI},N}=\sqrt{1+4\ta\cL_\gamma^{{\MS},N}+4\ta^2 \bP_N^2}\;.
    \label{eq:MSBINLag}
\end{equation}
Clearly $\cL_0^{{\MS},N}=\cL^{{\Sc},N}$ and $\cL_{\ta,0}^{{\MSBI},N}=\cL_\ta^{{\scriptscriptstyle{\text{NG}}},N}$ where $\cL_\ta^{{\scriptscriptstyle{\text{NG}}},N}$ is the Nambu-Goto Lagrangian density in a $d=N+2$ target space and imposing the static gauge condition. A simple computation shows that the components of the stress-energy tensor associated to \eqref{eq:MSBINLag} are
\begin{equation}
\label{eq:TMSBI2}
(\T_{\ta,\gamma}^{{\MSBI},N})_{ab}=a_{\ta,\gamma}^{{\MSBI},N}\eta_{ab}+b_{\ta,\gamma}^{{\MSBI},N}H_{N,ab}\;,
\end{equation}
with coefficients
\begin{align}
  &a_{\ta,\gamma}^{{\MSBI},N}=\frac{\bP_N^2}{\cS^{{\MSBI},N}}\left(2\ta+\frac{\sinh (\gamma )}{\sqrt{\bS_N^2-\bP_N^2}}\right)+\frac{1-\cS^{{\MSBI},N}}{2 \ta } \;,\notag \\ &b^{{\MSBI},N}_{\ta,\gamma}=\frac{1}{\cS^{{\MSBI},N}}\left(\cosh (\gamma )-\frac{\bS_N\sinh (\gamma )}{\sqrt{\bS_N^2-\bP_N^2}}\right)\;.
  \label{eq:TMSBI2coeff}
\end{align}
Using \eqref{eq:TMSBI2} and \eqref{eq:TMSBI2coeff}, it is not difficult to prove that $\cL_{\ta,\gamma}^{{\MSBI},N}$ belongs to the flow of the $\TbT$ operator with $\cL_{\gamma}^{{\MS},N}$ as initial condition at $\ta=0$, for any value of $\gamma$. Moreover, $\cL_{\ta,\gamma}^{{\MSBI},N}$ belongs to the flow of the marginal operator
\begin{equation}
\label{eq:O2marg}
    \widetilde{\cO}_\gamma^{[2]}:=-\sqrt{-\cO_\gamma^{[\frac12,2]}}=-\frac{1}{\sqrt{2}}\sqrt{\tr[\BT_\gamma^2]-\frac12 \tr[\BT_\gamma]^2}\;,
\end{equation}
with $\cL_{\ta}^{{\scriptscriptstyle{\text{NG}}},N}$ as initial condition at $\gamma=0$, for any value of $\ta$. 

Moving from Euclidean $(x^0,x^1)$ to complex coordinates $(z,\bar{z}):=(x^1+\mathbbm{i}x^0,x^1-\mathbbm{i}x^0)$ and introducing the components $(\T,\bar\T,\Theta)$ of the stress-energy tensor in the coordinates $(z,\bar z)$ which are related to $(\T_{00},\T_{11},\T_{01})$ in the coordinates $(x^0,x^1)$ according to the standard convention \cite{Zamolodchikov:2004ce}
\begin{equation}
\label{eq:TtoT}
\T = \frac{\pi}{2}\left(\T_{00}-\T_{11}+ 2 \mathbbm{i} \T_{01}\right) \;,\; \bar\T = \frac{\pi}{2}\left(\T_{00}-\T_{11}- 2 \mathbbm{i} \T_{01}\right)\;,\; \Theta=\frac{\pi}{2}\left(\T_{00}+\T_{11}\right)  \;,
\end{equation}
then \eqref{eq:O2marg} can be rewritten in an alternative way as
\begin{equation}
\label{eq:O2margcomplex}
    \widetilde{\cO}_\gamma^{[2]}= -\frac{1}{\pi} \sqrt{\T_\gamma\bar\T_\gamma}\;.
\end{equation}
We conclude this section by noticing that the previous results admit a natural generalisation to the case of $N$ scalar fields $\{\phi_i\}_{i\in\{1,\dots,N\}}$ interacting with a generic potential $V=V(\phi_1,\dots,\phi_N)$. Taking inspiration from \eqref{eq:MSNLag} and \eqref{eq:MSBINLag}, we consider the following Lagrangian density
\begin{equation}
\label{eq:MBNLag}
    \cL_\gamma^{{\MS},N,V}=\cL_\gamma^{{\MS},N}+V\;,
\end{equation}
and its Born-Infeld-like extension 
\begin{equation}
    \cL_{\ta,\gamma}^{{\MSBI},N,V}=\cL_{\bta,\gamma}^{{\MSBI},N}+\frac{V}{1-\ta V}\;,
\end{equation}
with $\bta=\ta\left(1-\ta V\right)$. 
It is a matter of a simple computation to show that $\cL_{\ta,\gamma}^{{\MSBI},N,V}$
belongs to the flow of the $\TbT$ operator with $\cL_\gamma^{{\MS},N,V}$ as initial condition at $\ta=0$, for any value of $\gamma$. Moreover, $\cL_{\ta,\gamma}^{{\MSBI},N,V}$ belongs to the flow of the marginal operator \eqref{eq:O2marg} with $\cL_{\bta}^{{\scriptscriptstyle{\text{NG}}},N}+\frac{V}{1-\ta V}$ as initial condition at $\gamma=0$, for any value of $\ta$.

Finally, notice that a perturbation of a CFT in $d=2$ with the square root of the $\TbT$ operator was introduced in \cite{Rodriguez:2021tcz} in the study of the relation between relativistic and ultra/non-relativistic conformal algebra 
(see also \cite{Bagchi:2022nvj} for further interesting results on this subject).

\section{Conclusions}
This paper discusses important geometric features of specific  $\TbT$-type deformations in arbitrary dimensions. Various aspects and open problems deserve further investigation. 

The first natural question is whether there exist physically acceptable theories, in $\D \ne 1,2,4$, whose stress-energy tensors fulfil the constraint (\ref{eq:Ddiag}). More generally, it would be important to find an exact formula for the deformed metric without imposing strong constraints on the stress-energy tensor eigenvalues. For example, a milder assumption would be the tracelessness of the stress-energy tensor in flat space, which drastically simplifies the perturbative series. Nevertheless, we were unable to obtain a closed expression for the deformed metric. In  $d>2$ and besides Born-Infeld nonlinear electrodynamics, the general properties of this  $\TbT$-induced  geometric deformation and its phenomenological features as a perturbation of classical field theory problems are unknown and certainly deserve some investigation. Moreover, it remains an important open question whether the simple expression for the truncated metric \eqref{eq:hgcgAbelian} could lead to some exact quantum result in $\D=4$, for example, the Casimir energy in specific geometries. 

A further possible line of research concerns the extension of the current setup to encompass the $\TbT$-like deformation in arbitrary dimensions introduced in \cite{Bonelli:2018kik} and the deformation of supersymmetric theories \cite{Baggio:2018rpv,Chang:2018dge,Chang:2019kiu}. Concerning the results of section \ref{sec:MMBIdimred}, it would be nice to understand the properties of the Modified Scalar theories, viewed as marginal deformations of free boson CFTs.  

Finally, let us mention that the article \cite{Ondo:2022zgf} about $T^2$ deformations of large $N$ holographic CFTs appeared a day after the first version of the current paper was available on ArXiv. The perturbing operator considered in \cite{Ondo:2022zgf} is of the form \eqref{eq:Od} with $r=\frac{1}{d-1}$.   The setup and methodologies adopted  in the two papers are similar  in spirit  but slightly different, and it may be very instructive to explore the eventual connection between them. Moreover, \cite{Ondo:2022zgf} motivated the search for a one-parameter extension of our initial results, which were restricted to the case $r=\frac{2}{d}$.

\medskip
\noindent{\bf Acknowledgments --}
We are especially grateful to Stefano Negro for useful discussions and collaboration at early stages of this work. We greatly thank Hossein Babaei-Aghbolagh for suggesting to us the possible existence of the marginal operator \eqref{eq:O2marg}. We also thank Aritra Banerjee, Marco Bill\`o, Andrea Cavagli\`a, Christian Ferko, Ferdinando Gliozzi, Yunfeng Jiang, Leonardo Santilli, Ozgur Sarioglu, Alessandro Sfondrini and Vasudev Shyam for useful discussions, suggestions and comments.
This project was partially supported by the INFN project SFT, the EU network GATIS+, NSF Award  PHY-1620628, and by the FCT Project PTDC/MAT-PUR/30234/2017 ``Irregular connections on algebraic curves and Quantum Field Theory''. R.C. is also supported by the FCT Investigator grant IF/00069/2015 ``A mathematical framework for
the ODE/IM correspondence''.

\medskip
\noindent{\bf Note added --}
On June $22^{\text{nd}}$ 2022, the day the second version of this work was made public on arXiv, the manuscript~\cite{Ferko:2022lol}  appeared. In \cite{Ferko:2022lol}, it is independently shown that the operator \eqref{eq:O2marg} generates the $\gamma$-flow of the Lagrangian \eqref{eq:MSBINLag},  the setup is slightly more general, but the main equations and conclusions are in total agreement with ours.

\appendix
\section{Deformed vierbein for abelian gauge theories in $d=4$}
\label{sec:appendix}
Let us start from the general solution \eqref{eq:gsol} for $d=4$ 
\begin{equation}
\label{eq:gebar}
    g_{\mu\nu}(\var;\var_0)=\left(\eta_{ab}+\left(\var-\var_0\right)\hT_{\var_0,ab}\right)e_\mu^a e_\nu^b \;.
\end{equation}
We look for the vierbein $\bar{e}_\mu^a:=\bar{e}_\mu^a(\var;\var_0)$ such that 
\begin{equation}
    g_{\mu\nu}(\var;\var_0)=\bar{e}_\mu^a\,\bar{e}_\nu^b\,\eta_{ab}\;.
\end{equation}
Then, $\hat{e}_\mu^a$ and $\check{e}_\mu^a$ that fulfil \eqref{eq:hgcgjac} are obtained from $\bar{e}_\mu^a$ as follows
\begin{equation}
\label{eq:ebarehec}
    \hat{e}_\mu^a(\ta)=\bar{e}_\mu^a(\ta_0;\ta)\;,\quad \check{e}_\mu^a(\ta)=\bar{e}_\mu^a(\ta;\ta_0)\;.
\end{equation}
First of all, we decompose the stress-energy tensor $\T_{\var_0,ab}$ in flat space as 
\begin{equation}
    \T_{\var_0,ab}=a^{(0)}_{\var_0}\eta_{ab}+a^{(1)}_{\var_0} F_{ab}^2+a^{(2)}_{\var_0}F_{ab}^4 \;,
\end{equation}
which implies that
\begin{align}
    \hT_{\var_0,ab}&=\frac12\eta_{ab}\tr[\BT_{\var_0}]-\T_{\var_0,ab}\notag\\
    &=\eta_{ab}\left(a^{(0)}_{\var_0}+\frac12a^{(1)}_{\var_0}\,\tr[\BF^2]+\frac12a^{(2)}_{\var_0}\,\tr[\BF^4]\right)-\left(a^{(1)}_{\var_0} F_{ab}^2+ a^{(2)}_{\var_0} F_{ab}^4\right)\;.
\end{align}
Thus \eqref{eq:gebar} becomes
\begin{align}
    g_{\mu\nu}(\var;\var_0)&=e_\mu^a\, e_\nu^b\,\eta_{ab}+\left(\var-\var_0\right)e_\mu^a\, e_\nu^b\Biggl(\eta_{ab}\left[a^{(0)}_{\var_0}+\frac12a^{(1)}_{\var_0}\,\tr[\BF^2]+\frac12a^{(2)}_{\var_0}\,\tr[\BF^4]\right]\notag\\
    &-a^{(1)}_{\var_0} F_{ab}^2- a^{(2)}_{\var_0} F_{ab}^4\Biggr) \;.
    \label{eq:gebar2}
\end{align}
We make the following ansatz for $\bar{e}_\mu^a$
\begin{equation}
\label{eq:ansatzebar}
   \bar{e}_\mu^a=\sqrt{\Sigma(\var;\var_0)}\,e_\mu^a+\sqrt{\var-\var_0}\left(u^{(1)}_{\var_0}F_{cb}+u^{(2)}_{\var_0}F^3_{cb}\right)\eta^{ac}e_\mu^b\;,
\end{equation}
where we defined
\begin{equation}
\label{eq:Sigmadef}
    \Sigma(\var;\var_0)=1+\left(\var-\var_0\right)\left(v^{(0)}_{\var_0}+v^{(1)}_{\var_0}\tr[\BF^2]+v^{(2)}_{\var_0}\tr[\BF^4]\right)\;,
\end{equation}
and the coefficients $\left(u^{(1)}_{\var_0},u^{(2)}_{\var_0},v^{(0)}_{\var_0},v^{(1)}_{\var_0},v^{(2)}_{\var_0}\right)$ must be fixed in terms of $\left(a^{(0)}_{\var_0},a^{(1)}_{\var_0},a^{(2)}_{\var_0}\right)$. Using the ansatz \eqref{eq:ansatzebar} and formula \eqref{eq:Sigmadef} together with the fact that $F_{ab}$ and $F^3_{ab}$ are anti-symmetric tensors, we have
\begin{align}
    \bar{e}_\mu^a\,\bar{e}_\nu^b\,\eta_{ab}&=\Sigma(\var;\var_0)\,e_\mu^a\,e_\nu^b\,\eta_{ab}-\left(\var-\var_0\right)\left[\bigl(u^{(1)}_{\var_0}\bigr)^2 F_{ab}^2+2u^{(1)}_{\var_0}u^{(2)}_{\var_0}F_{ab}^4+\bigl(u^{(2)}_{\var_0}\bigr)^2 F_{ab}^6\right]e_\mu^a\,e_\nu^b\notag\\
    &=e_\mu^a\,e_\nu^b\,\eta_{ab}+\left(\var-\var_0\right)e_\mu^a\,e_\nu^b\Biggl(-F_{ab}^2\left[\frac18\bigl(u^{(2)}_{\var_0}\bigr)^2\left(2\,\tr[\BF^4]-\tr[\BF^2]^2\right)+\bigl(u^{(1)}_{\var_0}\bigr)^2\right]\notag\\
    &+\eta_{ab}\left[v^{(0)}_{\var_0}+v^{(1)}_{\var_0}\,\tr[\BF^2]+v^{(2)}_{\var_0}\,\tr[\BF^4]\right]-F_{ab}^4\left[2u^{(1)}_{\var_0}u^{(2)}_{\var_0}+\frac12\bigl(u^{(2)}_{\var_0}\bigr)^2\,\tr[\BF^2]\right]\Biggr)\;,
    \label{eq:ebareta}
\end{align}
where in the last equality we used the Cayley-Hamilton Theorem to write
\begin{equation}
F_{ab}^6=\frac{1}{2} F_{ab}^4 \,\tr[\BF^2]+\frac{1}{8} F_{ab}^2 \left(2\,\tr[\BF^4]-\tr[\BF^2]^2\right)\;.
\end{equation}
Imposing the equivalence between the RHS of \eqref{eq:gebar2} and \eqref{eq:ebareta} we obtain
\begin{align}
    &u^{(1)}_{\var_0}=\frac{\sqrt{2}\left(a^{(1)}_{\var_0}+\frac{\cV_{\var_0}}{4}\right)}{\sqrt{4a^{(1)}_{\var_0}+\cV_{\var_0}+a^{(2)}_{\var_0}\tr[\BF^2]}} \;,\quad u^{(2)}_{\var_0}=\frac{\sqrt{2} a^{(2)}_{\var_0}}{\sqrt{4a^{(1)}_{\var_0}+\cV_{\var_0}+a^{(2)}_{\var_0}\tr[\BF^2]}}\;,\notag\\ 
    &v^{(0)}_{\var_0}=a^{(0)}_{\var_0}\;,\quad v^{(1)}_{\var_0}=\frac12 a^{(1)}_{\var_0} \;,\quad v^{(2)}_{\var_0}=\frac12 a^{(2)}_{\var_0}\;,
    \label{eq:uvrelations}
\end{align}
with
\begin{equation}
    \cV_{\var_0}=\sqrt{\bigl(4a^{(1)}_{\var_0}\bigr)^2+8 a^{(1)}_{\var_0} a^{(2)}_{\var_0} \tr[\BF^2]+2\bigl(a^{(2)}_{\var_0}\bigr)^2\left(\tr[\BF^2]^2-2 \tr[\BF^4]\right)}\;.
\end{equation}

\bibliography{Biblio3}

\end{document}